\documentclass[a4paper,aps,pre,twocolumn,groupedaddress,showkeys,showpacs,floats,floatats,floatfix]{revtex4-1} 
\usepackage[utf8]{inputenc}
\usepackage[english]{babel}
\usepackage{graphicx}
\usepackage{dcolumn} 
\usepackage{bm}
\usepackage{natbib}
\usepackage{latexsym}
\usepackage{mathrsfs}
\usepackage{amssymb}
\usepackage{amsmath}
\usepackage{amscd}
\usepackage{color}
\usepackage{pifont}
\usepackage{pstricks,pst-node,pst-text,pst-3d}
\usepackage{verbatim}
\usepackage[T1]{fontenc}

\begin{document}  

\title{Comment on ``Lyapunov statistics and mixing rates for intermittent systems"}

\author{Roberto Artuso${}^{1,2,}$ } \email{roberto.artuso@uninsubria.it}
\author{Cesar Manchein${}^{1,3,}$} \email{cmanchein@gmail.com}

\affiliation{${}^1$Center for Nonlinear and Complex Systems
and Dipartimento di Scienza e Alta Tecnologia \\ Via Valleggio 11, 22100 Como (Italy); \\
${}^2$I.N.F.N., Sezione di Milano, Via Celoria 16, 20133 Milano (Italy); \\
${}^3$Departamento de F\'isica, Universidade do Estado de Santa Catarina  89219-710 Joinville, (Brazil)}

\date{\today}

\begin{abstract}
In Pires {\it et al.} [Phys. Rev. E 84, 066210 (2011)] intermittent maps are considered, and the tight relationship between correlation decay of smooth observables and large deviations estimates, as for instance employed in Artuso and Manchein [Phys. Rev. E 80, 036210 (2009)], is questioned. We try to clarify the problem, and provide rigorous arguments and an analytic estimate that disprove the objections raised in Pires {\it et al.} [Phys. Rev. E 84, 066210 (2011)] when ergodic systems are considered. 
\end{abstract}


\maketitle

\section{Introduction}

In a recently published paper \cite{PSV}, the authors argue that a close relationship between polynomial large deviations and power law decay fails, in the form considered in \cite{AM}, which was inspired by the rigorous results of \cite{MelLM} (earlier results, which are relevant to the present discussion are \cite{ALP,PS}). To this end they both consider ergodic and infinite ergodic systems: the present comment concerns the first case, namely the original setting of \cite{AM}: we point out that the discussion in \cite{AM} concerned {\it mixing} systems with a positive Lyapunov exponent (see eq. (1) in \cite{AM}, where the relevant quantity is identically zero in the infinite ergodic regime). We also point out that the very concept of mixing is still
under debate in the framework of infinite ergodic theorem \cite{Len}, though some hints that correlation decay and large deviations are indeed related has recently appeared \cite{Gmix}. We regret that the notion of ``weak chaos" is loosely defined, even though any confusion between ergodic and infinite ergodic system was, in our opinion, {\it never} present, since an unambiguous definition of the main quantities employed in \cite{AM} is only possible in the positive Lyapunov - ergodic regime.
 
To keep the discussion as simple as possible we focus on the main dynamical example considered in \cite{PSV}, namely the intermittent Pomeau-Manneville map \cite{PomMan80}, indeed a prototype example of weak chaos (in any possible meaning of such a concept): this is defined on the unit interval $I=[0,1]$ and may be expressed as
\begin{equation}
\label{PMmap}
x_{n+1}=T(x_n)=\left. x_n + x_n^z \right|_{mod\,\,1},
\end{equation}
where, for $z>1$, the map presents in $x=0$ an indifferent fixed point, which influences deeply the dynamics (see, for instance, \cite{GW,Wang}). From the point of view of dynamical properties two regimes have to be considered: using the same terminology and notation of \cite{PSV}, we have an {\it exponential instability} regime, for $z \in (1,2)$, where an invariant probability measure $\mu$ exists ({\it i.e.} the invariant density is renormalizable), and where we have a Lyapunov exponent $\Lambda_\infty >0$, and a {\it subexponential instability} one, for $z>2$, where the invariant measure is not renormalizable, and no positive Lyapunov exponent exists: this is one of the prominent examples where {\it infinite ergodic theory} applies (see, for instance, \cite{Aar}).
In this comment we will consider the {\it exponential instability} regime: this is representative of a wide class of physically relevant systems where, together with exponential instability, long time tails for correlations are present: examples are Lorenz type maps \cite{CHMV}, area preserving maps \cite{CK}, Sinai \cite{bunJ}, stadium \cite{cgv} and mushroom \cite{bunC} billiards.
In the {\it exponential instability} case of Pomeau Manneville maps it is well known that ``generic" correlation functions decay with a well defined power-law exponent (we will later on comment on the precise meaning of ``generic" in this context): for example, by considering both lower and upper bounds, in \cite{hu} it is proved that there are Lipschitz functions $F$ and $G$ for which
\begin{equation}
\label{cor-hu}
\left| \int_I\, d\mu \, (F\circ T^n)G -\int_I\, d\mu \,F \, \int_I\, d\mu \,G \right|=O(n^{-(\frac{1}{z-1}-1)});
\end{equation}
while the upper bound holds for any pair of Lipschitz functions, the lower bound cannot share this generality, and indeed for the {\emph{non-generic}} case where $\int_I\,d\mu \, F=F(0)$ the decay is faster \cite{Gou} (it gains a $1/n$ factor \cite{nn}). We point out again that both in \cite{PSV} and \cite{AM} correlation decay are considered for sufficiently smooth functions: it is well known (see for instance \cite{cc} in the case of cat maps) that if one considers larger function spaces where test functions are picked from, one may get very different behavior. 

Now we have to consider large deviation, in order to see whether  their behavior is linked to correlation decay (as in \cite{AM}), or not, as claimed in \cite{PSV}. More precisely we consider (as in \cite{AM}) the distribution of finite time Lyapunov exponents
\begin{equation}
\label{ftle}
\Lambda_n(x)=\frac{1}{n}\sum_{j=0}^{n-1}\,\ln T'(T^j(x)),
\end{equation}
where the corresponding probability density is denoted by $\eta(\Lambda, n)$ \cite{n3}, and consider the quantity
\begin{equation}
\label{Mtail}
{\cal M}_{\tilde{\Lambda}}(n)=\int_0^{\tilde{\Lambda}}\,d \Lambda\, \eta(\Lambda,n),
\end{equation}
for $0 < \tilde{\Lambda} <\Lambda_{\infty}$: this of course implies that $\Lambda_{\infty}>0$, and so our considerations apply to the \emph{exponential instability} case, $1<z<2$ \cite{n1}. In \cite{AM} (on the grounds of \cite{MelLM,ALP}) it is stated that ${\cal M}_{\tilde{\Lambda}}(n)$ has the same polynomial behavior of {\emph{generic}} correlation functions (\emph{i.e} it vanishes according to a power law with the same exponent), while in the section $II.A$ of \cite{PSV} it is claimed that such a quantity decays {\emph{exponentially}}. 

This is the crucial point of this Comment: we now prove that the arguments of \cite{PSV} are incorrect. Their claim that ${\cal M}_{\tilde{\Lambda}}(n)$ decays exponentially consist in \emph{i.)} (their Eq. ($8$)) supposing that in such a case a good rate function exists for the observable $g(x)=\ln T'(x)$, and \emph{ii.)} that such a claim is consistent with a theorem proved by Pollicott and Sharp \cite{PS}. The subject of large deviations in the context of dynamical systems is indeed very important, and the existence of nice rate functions is not generically expected unless trajectories enjoy very good statistical properties \cite{LSY,LSY2}. In the present case, a number of rigorous results \cite{MN,PS,MelLM,AFLV}, that apply to the Pomeau-Manneville map in the \emph{exponential instability} case, concern exactly \emph{polynomial} large deviations. As it is crucial in the discussion let us recall the theorem proved in \cite{PS} (the formulation takes into account a sharper result obtained in \cite{MelLM}).  The theorem applies to any H\"older observable $g(x)$ with zero average $\int_I\,d\mu \, g=0$.
The theorem consists of two large deviation statements: if
$|g(0)|>\epsilon$ then
\begin{equation}
\label{PSpol}
\mu\left\{x \in I: \left|\frac{1}{n}\sum_{i=0}^{n-1}g(T^ix)\right| \geq \epsilon \right\}=O(n^{-(\frac{1}{z-1}-1)});
\end{equation}
(\emph{polynomial} case), while, if $|g(0)|< \epsilon$
\begin{equation}
\label{PSexp} 
\mu\left\{x \in I: \left|\frac{1}{n}\sum_{i=0}^{n-1}g(T^ix)\right| \geq \epsilon \right\}=O(e^{-\beta n}),
\end{equation} 
for some $\beta>0$
(\emph{exponential} case). Now the observable we have to consider is $\hat{g}(x)=\ln T'(x)$: while indeed $\hat{g}(0)=0$,
this function has a non-zero average $\int_I\, d\mu\,\hat{g}=\Lambda_{\infty}$, and so, to apply the theorem, we have to consider the observable $\breve{g}=\Lambda_{\infty}-\ln T'(x)$. Since $\breve{g}(0)=\Lambda_{\infty}$ the \emph{exponential} part of the theorem (\ref{PSexp}) (invoked in \cite{PSV}) provides no information about ${\cal M}_{\tilde{\Lambda}}(n)$ (it concerns $\eta(\Lambda,n)$ for $\Lambda <0$, which is identically zero), while the \emph{polynomial} case (\ref{PSpol}) gives exactly
\begin{equation}
\label{Mpol}
{\cal M}_{\tilde{\Lambda}}(n)=O(n^{-(\frac{1}{z-1}-1)}) \,\, \forall \tilde{\Lambda}\,\,\mathrm{such}\,\,\mathrm{that}\,\,0<\tilde{\Lambda}<\Lambda_{\infty},
\end{equation}
in perfect agreement with \cite{AM}, contrarily to what the authors of \cite{PSV} claim. Physically the mechanism that leads to the \emph{non-generic} exponential large deviations for intermittent dynamics (\ref{PSexp}) is very simple: while statistical anomalies are due to long waiting times near the indifferent fixed point, in the fine-tuned case where the value of the observable at the indifferent fixed point coincides with the phase average of the observable, during the laminar sequence Birkhoff sums pick up the right value, concealing the dynamical anomalies due to sticking. Let us further add that obviously, as regards finite time Lyapunov exponents, these anomalies manifest themselves in the distribution $\eta(\Lambda,n)$ for small $\Lambda$,  $0<\Lambda<\Lambda_{\infty}$, like in \cite{ALP}, whose results are misquoted in \cite{PSV} (see their eq. ($4$)) \cite{n2}.
Let us finally remark that, mathematically, results like eq. (\ref{PSpol}) are upper bounds, and further discussion is needed as regards their optimal character: this is indeed discussed in \cite{MelLM}, where a lower bound (of the same power-law form) is shown to hold for a special class of functions, exactly in the case of Pomeau-Manneville maps.

Though the mathematical misunderstandings in \cite{PSV} are already clear at this point, we can also add an explicit analytic computation that illustrates the way in which polynomial large deviations are attained for Pomeau-Manneville maps. We will obtain a lower bound, that again shows how polynomial large deviations estimates are optimal for such maps, with a well defined exponent. In particular we use the Gaspard-Wang piecewice linear approximation, discussed in detail in \cite{Wang}, to which we refer for full details. The approximation consists in partitioning $I$ by the collection of sets $A_n=(\xi_{n},\xi_{n-1})$, where $\xi_{-1}=1$, $\xi_0=a$ (where $a$ is such that $T(a)=a+a^z=1$), and $\{\xi_n\}$ is the decreasing sequence converging to zero, such that $T(\xi_{n})=\xi_{n-1}$. Then the new (piecewise linear) map $T_L$ is constructed such that its slope is constant on every $A_n$, and $T_L(A_n)=A_{n-1}$. We denote by $\Delta_n=\xi_{n-1}-\xi_n$ the width of $A_n$ and by $s_n=\Delta_{n-1}/\Delta_n$ the slope of $\left. T_L \right|_{A_n}$. The invariant measure density is piecewise constant, and can be easily evaluated by considering the equivalent Markov chain, the invariant measure weights are $\mu_n=\mu(A_n)$, and again in the case $z \in (1,2)$ we have an invariant \emph{probability} measure, which shares all the relevant properties of the corresponding Pomeau-Manneville invariant \emph{probability} measure for the same exponent $z$ \cite{n4}, moreover the correlation decay rate of \emph{generic} smooth functions is identical to (\ref{cor-hu}) \cite{Is}. For large $n$ we can easily get the dominant behavior \cite{Wang}:
\begin{equation}
\label{plEst}
\Delta_n \sim (n+1)^{-\frac{z}{z-1}},\qquad \mu_n\sim n^{-\frac{1}{z-1}}.
\end{equation} 
Now we define intervals near the fixed point ${\cal A}_k= \cup_{m=k}^\infty \, A_m$; for each $y \in {\cal A}_{n+N}$ the maximal finite time Lyapunov exponent we may obtain is
\begin{equation}
\label{maxLn}
\Lambda_{max}^{(N)}(n)=\frac{1}{n}\sum_{k=N+1}^{N+n}\ln s_k=\frac{1}{n} \ln \left( \frac{\Delta_{N}}{\Delta_{N+n}}\right)
\end{equation}
which, for sufficiently high $N$, has the following behavior (see (\ref{plEst})):
\begin{equation}
\label{max-est}
\Lambda_{max}^{(N)}(n)\sim \frac{1}{N+1}\left(\frac{N+1}{n} \right) \ln \left(1+\frac{n}{N+1} \right).
\end{equation}
Since, away from zero, the function $(1/y) \ln (1+y)$ is bounded, (\ref{max-est}) implies that, for any $\tilde{\Lambda}$ with $0<\tilde{\Lambda}<\Lambda_\infty$ we can fix a value $\tilde{N}$ such that $\Lambda_{max}^{(\tilde{N})}(n) < \tilde{\Lambda}$ for any $n$. This leads to the following bound:
\begin{equation}
\label{LDbou}
\mu\left\{x\in I :  \frac{1}{n} \sum_{k=0}^{n-1} \ln T_L'(T^k_Lx)< \tilde{\Lambda} \right\} \geq \mu({\cal A}_{n+\tilde{N}});
\end{equation}
now, by using (\ref{plEst}), we easily get
\begin{equation}
\label{fin-est}
\mu({\cal A}_{n+\tilde{N}})=\sum_{m=n+\tilde{N}}^\infty \, \mu(A_n) \sim (n+\tilde{N})^{-(\frac{1}{z-1}-1)}:
\end{equation}
\emph{i.e. polynomial} large deviations with the \emph{same} exponent ruling correlation decay. Though we used the piecewise linear approximation, the same kind of asymptotics on interval scaling and measures are known to hold for the original map (see for instance \cite{hu,Gou}).

This work has been partially supported by MIUR--PRIN 2008 project {\em  Nonlinearity and disorder in classical and quantum transport processes}.

\end{document}